%% file: disorderedopenfreefermions.tex
\documentclass[%
 reprint,
 amsmath,amssymb,
 aps,
]{revtex4-2}

\usepackage{cancel}
\usepackage{lipsum, babel}
\usepackage{graphicx}
\usepackage{dcolumn}
\usepackage{bm}
\usepackage{float}
\usepackage{soul}
\usepackage[colorlinks = true, citecolor = blue, linkcolor = blue, urlcolor=blue]{hyperref}
\usepackage{xcolor}
\usepackage{braket}

\newcommand{\tr}{\mathrm{Tr}}

\newcommand{\highlightchange}[1]{#1}
\newcommand{\highlightdelete}[1]{}
\newcommand{\removeblock}[1]{}

\begin{document}

\preprint{APS/123-QED}

\title{Entanglement and localization in long-range quadratic Lindbladians}

\author{Alejandro Cros Carrillo de Albornoz}
\affiliation{Department of Physics and Astronomy, University College London, Gower Street, London WC1E 6BT, United Kingdom}

\author{Dominic C. Rose}
\affiliation{Department of Physics and Astronomy, University College London, Gower Street, London WC1E 6BT, United Kingdom}

\author{Arijeet Pal}
\affiliation{Department of Physics and Astronomy, University College London, Gower Street, London WC1E 6BT, United Kingdom}

\date{\today}

\begin{abstract}

Existence of Anderson localization is considered a manifestation of coherence of classical and quantum waves in disordered systems. Signatures of localization have been observed in condensed matter and cold atomic systems where the coupling to the environment can be significantly suppressed but not eliminated. In this work we explore the phenomena of localization in random Lindbladian dynamics describing open quantum systems. We propose a model of one-dimensional chain of non-interacting, spinless fermions coupled to a local ensemble of baths. The jump operator mediating the interaction with the bath linked to each site has a power-law tail with an exponent $p$. We show that the steady state of the system undergoes a localization entanglement phase transition by tuning $p$ which remains stable in the presence of coherent hopping.  Unlike the entanglement transition in the quantum trajectories of open systems, this transition is exhibited by the averaged steady state density matrix of the Lindbladian. The steady state in the localized phase is characterized by a heterogeneity in local population imbalance, while the jump operators exhibit a constant participation ratio of the sites they affect. Our work provides a novel realization of localization physics in open quantum systems.

\end{abstract}

\maketitle

\textbf{\textit{Introduction.---}} The interplay of coherence and disorder in quantum systems leads to the existence of the phenomena of localization \cite{EverMirlin_RMP2008}, shown to remain stable in the presence of many-body interactions \cite{Basko2006, Gornyi2005, Pal2010, Huse2007, Nandkishore_2015, Abanin2019}. 
Such localized systems protect coherence and provide a route to stabilizing quantum phases of matter out of equilibrium \cite{Huse2013, Bahri2013, Parameswaran2018}.  
Even though localization is stable in isolated quantum systems, the presence of interactions with a bath can destroy the coherence \cite{Levi2016, Medvedyeva2016, Luschen2017, Lenar2018, Lunt2020, Lenar2020, Wybo2020, Hamazaki2022, kawabata2022SYK, garcia2022SYK, szyniszewski2022disorderFF, mak2023NHMBL}. 
In particular, the role of baths in destabilizing many-body localization resulting in a thermal steady state is of fundamental importance \cite{Sels_deloc_PRB2022, Morningstar2022}. 
Interaction with baths are unavoidable for experiments probing late time behavior, and a theoretical understanding is required for the interpretation of the observations \cite{Luschen2017, Fitzpatrick2017, JiamingLi2019PT, XiaoMi_DTC2022}  

Open quantum systems (OQS) can host steady states which are manifestly out of equilibrium and protect coherence from environmental noise, such as decoherence free subspaces \cite{Zanardi1997a,Zanardi1997b,Lidar1998} and
noiseless subsystems \cite{Knill2000,Zanardi2000}, among other non-equilibrium phenomena \cite{Albert2016,Landi2021,Sieberer2016,Maghrebi2016,Torre2013}.
The long time dynamics leading up to the stationary state have also been shown to present complex metastable behavior \cite{Macieszczak2016,Rose2016,Macieszczak2021,Rose2022}, while periodic driving can enable time crystalline phenomena \cite{Iemini2018,Zhu2019,Gambetta2019,RieraCampeny2020, Sullivan_dDTC_NJP2020, Kessler_dTC_PRL2021,Fazio2022,Buca2019}. 
Quantum trajectories of OQS, providing individual stochastic realizations of the systems pure-state evolution, can exhibit non-trivial trajectories statistics that cannot be detected in either the averaged time-evolution of the density matrix or its steady state \cite{Garrahan2010,Carollo2018,Carollo2019,Carollo2020,Carollo2021,Olmos2012,Olmos2014,Lesanovsky2013,Marcantoni2021,Hickey2012}.
In particular, recent work demonstrating entanglement transitions in quantum trajectory ensembles has spurred interest for quantum information processing \cite{liMeasurementdrivenEntanglementTransition2019, skinnerMeasurementInducedPhaseTransitions2019, liQuantumZenoEffect2018}.
Altogether, the preservation of coherence in non-equilibrium OQS suggests that it may also be possible for localized behavior to survive in the presence of dissipation.

\begin{figure}[h!]
    \centering
    \includegraphics[width=\columnwidth]{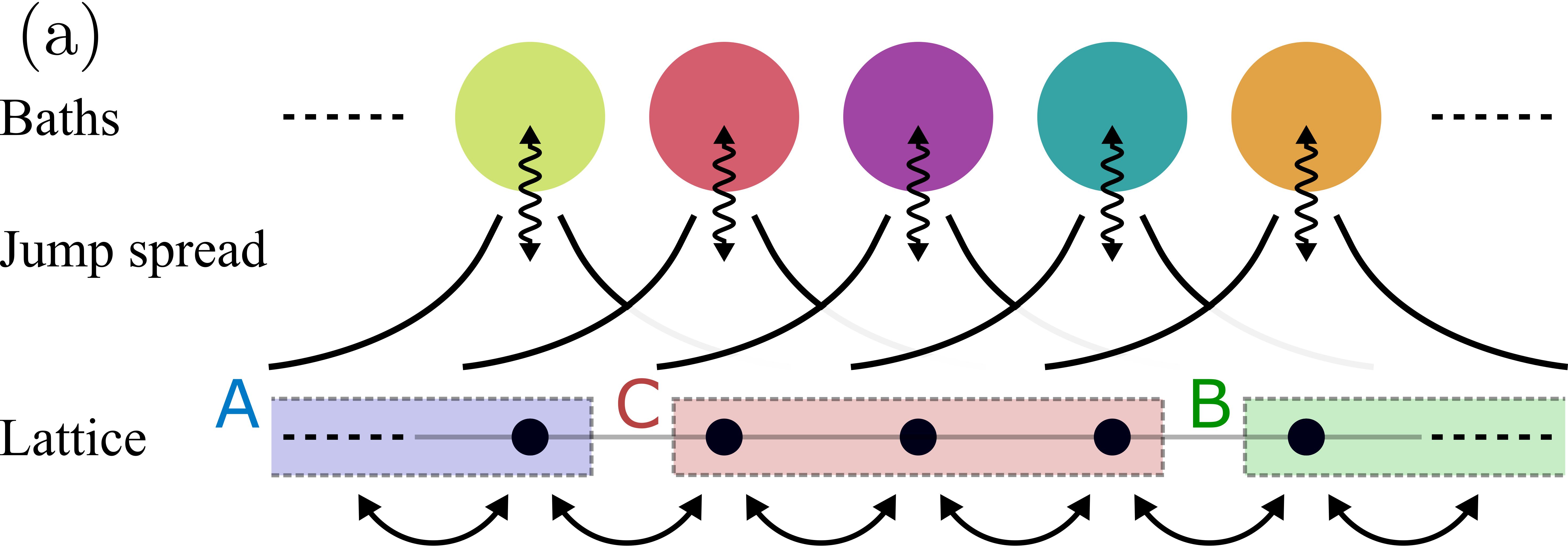}
    \includegraphics[width=\columnwidth]{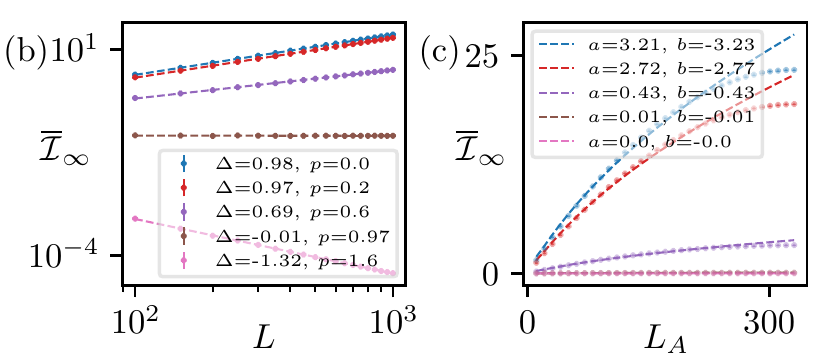}
    \caption{
    (a) Schematic figure of the model: The spatial power-law profile of each jump operator, given by exponent $p$, for the heterogeneous baths shows the coupling to a system of spinless fermions hopping in a 1D lattice. 
    We consider the mutual information between fermions in region A and B ($\mathcal{I}$). 
    (b) Disorder averaged mutual information between $A$ and $B$ in the steady state ($\overline{\mathcal{I}}_\infty$) as a function of system size $L$, with $L_A=L_B=L_C=L/3$ 
    For each value of $p$ (shown in the legend) the curves are fitted to the form $\ln(\overline{\mathcal{I}}_\infty)=\Delta\ln(L)+c$. The fitting parameter $\Delta$ is shown in the legend.
    (c) $\overline{\mathcal{I}}_\infty$ between $A$ and $B$ in the steady state as a function of subsystem size $L_A$ with total system size $L=10^3$. $L_C$ is fixed at $L/3$. Curves were fitted according to $\overline{\mathcal{I}}_\infty = a(L_A)^c + b\log(L_A)$. Note $c\approx 1/2$ for all $p$.
    For both (b), (c): Curves are color-coded to a given $p$ according to the legend in panel (b).
    Error bars are not visible. 
    Details on disorder realizations are included in SM7.
    }
    \label{fig:cartoon-and-MI}
\end{figure}

Quadratic Lindbladians have emerged as a testbed for studying various concepts in OQS \cite{Prosen2008a,Prosen2008b,Guo2017,Prosen2010,Mahajan2016,Zanoci2016,Lieu2020,Barthel2022,DAbbruzzo2022,Costa2022, Vernier2020}, due to simplifications enabled by their non-interacting nature.
In this article we propose and investigate the properties of a quadratic fermionic model of localization induced by coupling to an heterogeneous bath, where the local degrees of freedom couple to randomly varying baths in space, similar to recent work on random Lindbladians \cite{Denisov2019,Can2019a,Can2019b,Sa2020a,Sa2020b,Wang2020,Sommer2021,Tarnowski2021,Lange2021,Li2022,Sarang2022}. 
The baths are modeled using a Lindbladian formalism for OQS where the jump operators couple distant sites, realizable in the collective dissipation of atomic arrays \cite{Yamamoto2016,Marino2022,Seetharam2022,Norcia2018,Miranda2017,Needham2019,Jones2018,Wang2021}.
The locality of the jump operators mediating the interaction via the bath is a tunable parameter, driving a phase transition between localized and delocalized steady states, the strength of the couplings decaying as a power-law in separation between sites in the localized phase.
The steady state properties closely relate to the single-particle localization transition of power-law banded matrices for purely unitary dynamics  \cite{Fyodorov1991,Fyodorov1996,Casati1990,Casati1993,Schenker2009,Bourgade2018,Varga2002,Cao2017}.
Furthermore, we studied the effect of introducing a coherent hopping term on the system's dynamics and steady state, finding the phenomena observed - including the transition -  are largely independent of such perturbations.

\begin{figure}[h!]
    \centering
    \includegraphics[width=\columnwidth]{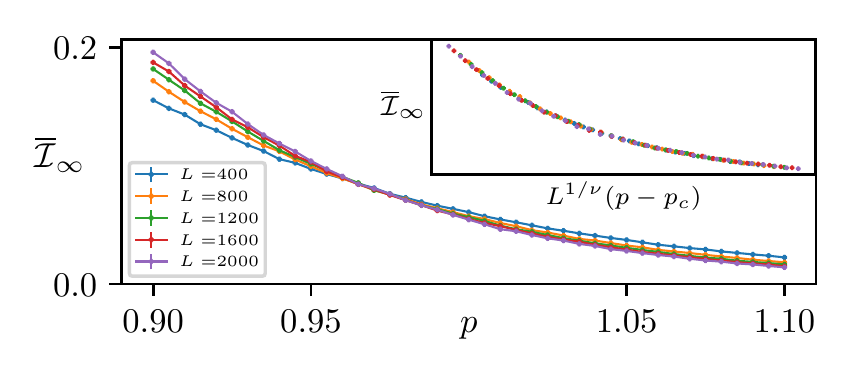}
    \caption{
    Mutual information between $A$ and $B$ of the system as a function of $p$, with $L_A=L_B=L_C=L/3$, for a set of system sizes $L$ shown in the legend, averaged over disorder. Inset shows a finite size scaling collapse with $p_c=0.98 \pm 0.01$ and $\nu=6.3 \pm 0.7$. Error bars are not visible. Details on disorder realizations and fitting procedure are included in SM7.
    }
    
    \label{fig:MI-finite-size-scaling}
\end{figure}

\noindent \textbf{\textit{Model.---}}We consider a system of one-dimensional spin-less fermions modeled by a quadratic Lindbladian, where the Hamiltonian conserves particle-number, and with spatially heterogeneous dissipative terms (see SM1 for details). 
The quadratic structure of Lindbladian can be exploited to derive an equation for the time evolution of the two point correlation matrix $\Omega_{jk} = \textrm{Tr}(\rho c^\dagger_j c_k)$, where $c_i^{\dagger}$ is a fermionic creation operator at site $i$  \cite{Mahajan2016}. We choose pairs of jump operators which add or remove fermions with rates that sum to one, leading to
\begin{align}\label{eq:correlator-matrix-diff-eq-simplified}
    \frac{d\Omega}{dt} 
    &= i[h^T,\Omega]+\frac{1}{2}\left(\{\Gamma,\mathbb{I}-\Omega\}-\{\mathbb{I}-\Gamma,\Omega\}\right),
\end{align}
where $h$ is a hermitian matrix \highlightchange{accounting for} \highlightdelete{representing the contribution of} the Hamiltonian, and $\Gamma$ \highlightchange{models} \highlightdelete{is modeling the} bath interactions.
\highlightchange{We consider cases where $h$ is either $0$, or represents coherent hopping with $h_{nm} = \lambda$ if $m = n\pm1$ and $0$ otherwise.
However, note Eq. \eqref{eq: dynamical solution no H} and Eq. \eqref{eq: general solution to omega} hold for any $h$.}
The second term of Eq.~\ref{eq:correlator-matrix-diff-eq-simplified} corresponds to bath interactions which create fermions within the system, while the third term
removes them.
To maintain hermiticity and positivity of $\Gamma$ we take $\Gamma\propto X^\dagger X$ for a random matrix X.
The elements of $X$ are chosen as
\begin{equation}\label{eq: the model}
    X_{jk} = \frac{x_{jk}}{(|j-k|+1)^p},
\end{equation}
where $x_{jk}$ is chosen from the complex Gaussian distribution $\mathcal{C}\mathcal{N}(0,1)$, and $p$ controls the decay of matrix elements with distance from the diagonal, in close analogy with random power-law banded and Wishart matrices.
This corresponds to matrix elements decaying with distance from each site in a system with open boundary conditions. \highlightchange{Long-range jump operators with a power-law spatial profile can be realized in cold atoms in an optical cavity with tunable Raman side bands for the driving field discussed in~\cite{Seetharam2022}.}
To maintain the validity of the dissipative evolution given by Eq.~\ref{eq:correlator-matrix-diff-eq-simplified} the maximum eigenvalue of $\Gamma$ must be less than $1$, so we scale the product by twice its maximum eigenvalue, which we denote $\lambda_\textrm{max}$, i.e. $\Gamma=X^\dagger X / 2\lambda_\textrm{max}$. \highlightchange{For an alternative approach see the Kac normalisation in Ref.~\cite{Passarelli_2022}}. The overall scaling does not effect the local properties or the steady state, which will be the focus of our work.

The value of $p$ directly influences how localized the eigenvectors of $\Gamma$ are, leading to the model sketched in Fig.~\ref{fig:cartoon-and-MI}\textcolor{blue}{(a)}.
Each site can be imagined to have an associated bath, interacting with the system with distinct rates which can be viewed as representing differing energies of the modes or different temperatures of the baths, encoded in the eigenvalues of $\Gamma$.
These interactions are mediated via jump operators which add or remove particles from modes which are focused on the associated site, decaying away from that site as a power-law with an exponent $p$: these jump operators correspond to superpositions of single site creation or annihilation operators with coefficients given by the eigenvectors of $\Gamma$.

\noindent \textbf{\textit{Pure dissipation solution.---}}
The steady state, and time evolution leading to it, can be solved exactly both with and without a Hamiltonian.
For simplicity we begin with the Hamiltonian free case ($h=0$), finding
\begin{equation}\label{eq: dynamical solution no H}
    \Omega(t) = \Gamma(1-e^{-t})+\Omega(0)e^{-t}.
\end{equation}
The steady state correlation matrix is independent of the initial conditions and is therefore unique, equal exactly to $\Gamma$ in this case.
The parameter $p$ therefore allows us to tune between two limits.
When $p$ is large, implying $\Gamma$ is short ranged and almost diagonal, then the steady state is approximately unentangled. 
On the other hand, when $p=0$ it implies $\Gamma$ is infinite-ranged and random, and one would expect the steady state to be entangled and obey a volume law instead \cite{Magan2016,Tao2012}. 
In this work we study what happens between the two limits as $p$ is varied.

\noindent \textbf{\textit{Mutual information.---}} 
We first consider the \highlightchange{correlations} \highlightdelete{entanglement} of the steady state as we vary $p$ using the mutual information $\mathcal{I}$ (denoted by $\mathcal{I}_\infty$, with disorder average denoted by a bar $\overline{\mathcal{I}}_\infty$), defined between subsystems $A$ and $B$ as 
\begin{equation}\label{eq: mi def}
    \mathcal{I}_{A|B}(\rho)= S_A(\rho)+S_B(\rho)-S_{A\cup B}(\rho),
\end{equation}
where $S_X(\rho)$ is the von Neumann entropy of $\rho$ in subsystem $X$, which may be rewritten in terms of the subsystem correlation matrix $\Omega_X$ of $\rho$ as \cite{Peschel2009, Surace2022}
\begin{align}
    S_X(\rho)=-\tr\left[\Omega_X\ln(\Omega_X)+(1-\Omega_X)\ln(1-\Omega_X)\right].
\end{align}

We choose subsystems $A$ and $B$ separated by an intervening region $C$ of length $L_C=L/3$ to remove boundary \highlightchange{correlation} \highlightdelete{entanglement} terms, c.f. Fig.\ref{fig:cartoon-and-MI}\textcolor{blue}{(a)}, and consider two cases: varying overall system size $L$ with $L_A=L_B=L/3$; and varying subsystem size $L_A$ for a fixed $L$ 
\footnote{If the boundary terms are not removed the critical point is shifted to a slightly larger value. This is shown in SM6, where we consider the behavior of $\overline{\mathcal{I}}_\infty$ in a setup with boundary effects.}.
In the first case in Fig.~\ref{fig:cartoon-and-MI}\textcolor{blue}{(b)} we observe that for small $p$, $\overline{\mathcal{I}}_\infty$ scales as a volume law in $L$, as expected, while for large p the lack of boundary terms cause $\overline{\mathcal{I}}_\infty$ to decay towards zero with increasing $L$.
At intermediate values of $p$ we observe power-law behavior, with exponent $\Delta$ decreasing as $p$ increases until a critical point $p=p_c$ at which $\overline{\mathcal{I}}_\infty$ is approximately constant as a function of $L$.
As $p$ increases further, $\Delta$ becomes negative.
This dependence of $\Delta$ on $p$ is suggestive of a transition between localized and delocalized phases.

In Fig.~\ref{fig:MI-finite-size-scaling} we perform a scaling collapse assuming a continuous transition finding a critical point at $p_c=0.98 \pm 0.01$ with a correlation length critical exponent of $\nu=6.3 \pm 0.7$. As expected the critical point is close to the localization transition for \highlightchange{power-law random banded matrices} (PRBM) of $p_c\sim1$. Our analysis of the \highlightchange{correlations} \highlightdelete{entanglement} in the steady state differs from the existing studies of entanglement in eigenstates of PRBM models. However, the delocalized phase close to the transition exhibits a sub-volume law \highlightchange{correlations} \highlightdelete{entanglement} in analogy with the eigenstate \highlightchange{entanglement} transition of PRBM models \cite{Jia_PRB2008, Lydzba_PRL2020, Tomasi_PRL2020}. Interestingly, The critical exponent $\nu$ is comparable in size to the exponents for measurement induced transitions for quantum trajectories studied in long-range Clifford circuits and free-fermion Hamiltonians~\cite{Block2022,Minato2022,Muller2022,Shraddha2022}. 

In Fig.~\ref{fig:cartoon-and-MI}\textcolor{blue}{(c)} we see that with $L_C$ fixed at $L/3$, $\overline{\mathcal{I}}_\infty$ is independent of subsystem size $L_A$ in the localized phase at large $p$. 
In contrast, the delocalized phase at small $p$ exhibits a sub-extensive scaling of $\overline{\mathcal{I}}_\infty$, differing from the linear dependence seen in volume-law entangled pure states. 
Our results suggest that even for $p \sim 0$, when $\overline{\mathcal{I}}_{\infty}$ scales as a volume law with system size, the scaling with $L_A$ is subextensive. 
A possible explanation could be that the steady states, although delocalized, are only weakly entangled locally. 
This would be analogous to delocalized, non-ergodic states discussed in the context of Anderson localization on Bethe lattices \cite{DeLucaPRL2014, parisi2019BetheAnderson}.

\begin{figure}
    \centering
    \includegraphics[width=\columnwidth]{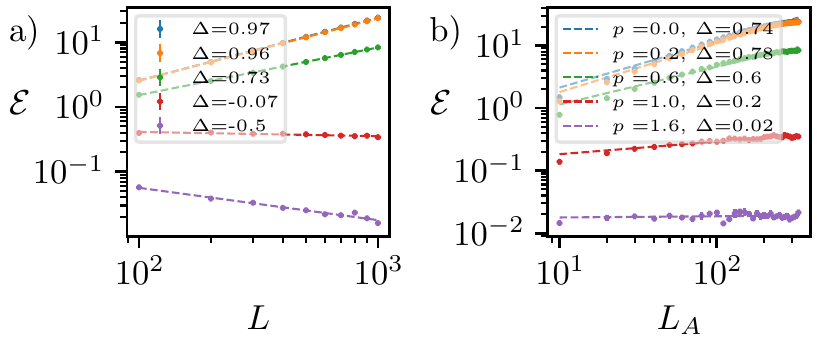}
    \caption{\highlightchange{(a) Disorder averaged entanglement negativity ($\mathcal{E}$) between $A$ and $B$ (compare with Fig.~\ref{fig:cartoon-and-MI}) in the steady state as a function of system size $L$, with $L_A=L_B=L_C=L/3$. (b) Steady state $\mathcal{E}$ between $A$ and $B$ as a function of subsystem size $L_A$ with $L_C=L/3$ fixed at $L=10^3$.
   For each value of $p$ (shown in the legend) the curves are fitted to the form $\ln(f(x))=\Delta\ln(x)+c$. The fitting parameter $\Delta$ is shown in the legend. Details on disorder realizations are included in SM7.
    }}
    \label{fig:negativity}
\end{figure}

\noindent \textbf{\textit{Entanglement negativity.---}} 
\highlightchange{
Since mutual information contains contributions from classical correlations, in order to quantify the quantum correlations in the two phases, we evaluate the entanglement negativity $\mathcal{E}$ of the steady state. Entanglement negativity is a bipartite entanglement measure of a mixed state defined through the positivity of the partially transposed density matrix \cite{vidal2002, Horodecki2000}. Computing the entanglement negativity can be hard even for Gaussian fermions as the partially transposed density matrix $\rho^{T_A}$ is not a Gaussian operator but a sum of two Gaussian operators $O_-$ and $O_+$ \cite{Zimboras_2015}. However, a similar entanglement monotone, the \textit{fermionic negativity} $\mathcal{E} = \ln \tr \sqrt{O_+ O_-}$, can be re-expressed as
\begin{equation}\label{eq: negativity}
    \mathcal{E} = \sum_j\ln\left[ \sqrt{\mu_j} + \sqrt{1-\mu_j} \right] + \sum_j\frac{1}{2}\ln\left[ (\lambda_j)^2 + (1-\lambda_j)^2 \right]
\end{equation}
where $\mu_j$ and $\lambda_j$ are eigenvalues of two algebraic expressions of the two-point correlator $\Omega$ \cite{alba2022logarithmic}.
In Fig.~\ref{fig:negativity} we show the variation of $\mathcal{E}$ with system size ($L$) and subsystem size ($L_A$). \highlightchange{We note that while $\mathcal{I}(L_A)$ (c.f. Fig.~\ref{fig:cartoon-and-MI}) suffers from finite size effects which become apparent in a log-log scale -- all curves scale in the same manner albeit the magnitude of $\mathcal{I}$ in the area law is $\mathcal{O}\left(10^{-7}\right)$ -- these finite size effects vanish in $\mathcal{E}(L_A)$, suggesting they are the result of the mixed nature of the steady state.}
}

\noindent \textbf{\textit{\highlightchange{Bath structure}.---}} 
\highlightdelete{To understand more about these phases, and the region around the phase transition,}
\highlightchange{The entanglement transition occurs in parallel with structural changes in the physical action of the bath. Due to the presence of the stationary state correlation matrix $\Gamma$ in the Lindblad equation (see SM1), its eigenstates inform us about the structure of the bath. Specifically, diagonalizing $\Gamma$, $\sum_k \Gamma_{jk}\phi^n_k = \gamma_n \phi^n_j$, allows writing the Lindbladian in the standard form using jump operators $d_n = \sum_j \phi^n_j c_j$. We thus consider the participation ratio (PR) of the eigenstates $\phi^n$ of $\Gamma$, a measure of their locality, defined as
\begin{align}
    \textrm{PR}(\phi)=\frac{1}{\sum_i|\phi_i|^4}.
\end{align}
If $\phi^n$ is very localised, then $d_n \sim c_n$ and the bath is onsite, whereas if $\phi^n$ is very delocalised then $d_n \sim \sum_n c_n$ and the bath couples all sites together regardless of their distance apart.}
\highlightdelete{The eigenstates inform us about the structure of the jump operators defining our open dynamics, and gives a sense of the locality of correlations within the steady state. For a general eigenvector $\phi$ we define the PR as}

\begin{figure}
    \centering
    \includegraphics[width=\columnwidth]{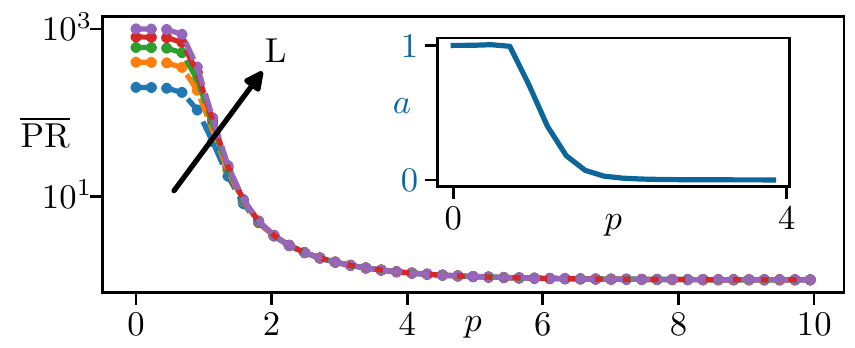}
    \caption{
    The disorder averaged participation ratio $\overline{\textrm{PR}}$ against $p$ for sizes $L=[400, 800, 1200, 1600, 2000]$, increasing along the arrow. 
    Inset: power $a$ as a function of $p$ found when fitting $\overline{\mathrm{PR}}(L,p)=L^{a(p)}/c(p)$ to $\overline{\textrm{PR}}(L)$ curves. 
    Data for $c$ is shown in SM4.
    }
    \label{fig:participation_ratio}
\end{figure}

In Fig.~\ref{fig:participation_ratio} we show $\overline{\textrm{PR}}$, the $\textrm{PR}$ averaged over eigenvectors of $\Gamma$  from multiple disorder realizations.
As $p$ is varied we see a transition from a delocalized phase where the $\overline{\textrm{PR}}$ grows linearly with $L$, to a region where it becomes small and constant.
However, rather than a sharp transition to a localized phase, we see a gradual reduction in the constant value that the PR achieves as $p$ increase, approximately reaching its lower bound of $1$ at large $p$.
This is exemplified in the inset of Fig.~\ref{fig:participation_ratio}\textcolor{blue}{(a)}, where we show the $p$ dependence of a power-law fit of $\overline{\textrm{PR}}(L)$.
We see the rapid decay of the exponent as $p$ is increased from the suspected critical point of around $p=1$, from the expected value of $1$ in the delocalized phase down to $0$ in the localized phase.

A possible explanation for the slow decay of the PR may be found in the localization behavior of the eigenstate.
In SM5.3 we observe that the eigenstates of $\Gamma$ exhibit a power-law decay away from some central site as found in prior studies of PRBM models \cite{Yeung1987}.
This stands in contrast to the exponential decay common in short-range models exhibiting localization.

\highlightchange{A related signature of long range structure is} multifractality, observed in the eigenstates of PRBM models near the transition \cite{Cuevas2001,Cuevas2003,Mace2019}.
A method of detecting this is to study different moments of probability distributions over subsystems derived from the systems eigenstates, specifically the scaling dependence of these moments vs the size of those subsystems.
At a transition, these exponents would be expected to be constant against system size; if the system exhibits multifractality, these exponents will take distinct values for different moments.
We verify that both these properties hold \highlightchange{in the eigenstates of $\Gamma$}, with the exponents characterizing this multifractality possessing values similar to prior PRBM work.
Details are provided in SM3.

\begin{figure}
    \centering
    \includegraphics[width=\columnwidth]{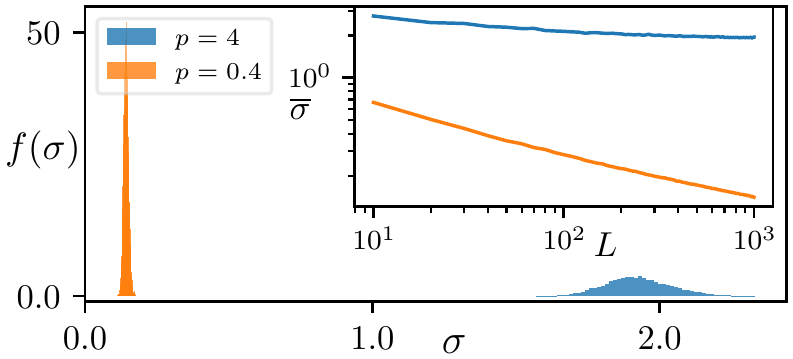}
    \caption{
    \highlightchange{Distribution of the standard deviation of the bias $\tilde{n}$ calculated for each realization using the central $L/5$ sites in the bulk in the localized (blue, $p=4)$) and delocalized (orange, $p=0.4$) phases for system size $L=10^3$. }
    Inset: average standard deviation as a function of $L$, error bars are not visible.}
    \label{fig:heterogeneity}
\end{figure}

\noindent \textbf{\textit{Stationary state heterogeneity.---}}
The presence of localization is usually accompanied by a lack of transport throughout the system, leading to \highlightchange{spatial} heterogeneity of physical quantities.
To observe such behavior in our system, we study the variability of single-site density matrices in the steady state. 
Mathematically, we may equate a single site density matrix to a diagonal matrix due to gaussianity. 
We write
\begin{align}
    \rho_s=\frac{e^{-\tilde{n}}\ket{o}\bra{o}+\ket{u}\bra{u}}{e^{-\tilde{n}}+1},
\end{align}
where $\ket{o}$ and $\ket{u}$ correspond to the single site $s$ being occupied and unoccupied, and $\tilde{n}$ is the bias of that site towards being unoccupied.
We study the distribution of $\tilde{n}$ for $p=0.4$ and $p=4$.
Figure \ref{fig:heterogeneity} shows the \highlightchange{standard deviation $\sigma$ distribution of $\tilde{n}$ from the sites within the middle $1/5$ of the chain calculated in multiple disorder} realizations. 
For this model $\tilde{n}$ is positive due to the spectra of $\Gamma$ being in the interval $[0, 1/2]$, resulting in a lower rate of particle creation than annihilation which biases the sites to be unoccupied. 
In the delocalized phase ($p=0.4$), the \highlightchange{standard deviation}\highlightdelete{bias} distribution is sharp and well defined \highlightchange{at small values} \highlightdelete{: characteristic of a system in thermal equilibrium.
In contrast, } whereas the localized phase ($p=4$) exhibits a distribution that is broad \highlightchange{at much larger values}\highlightdelete{and non-uniform: a signature of a heterogeneous system.}
\highlightchange{This suggests the localized phase is far more spatially heterogeneous within each realization, in addition to realization-to-realization variance.}
The inset of Fig.~\ref{fig:heterogeneity} further shows the scaling of the average standard deviation of $\tilde{n}$ against system size $L$, which is \highlightchange{weak} in the localized phase but \highlightchange{quickly} sharpens in the delocalized phase as system size increases, implying a homogeneous steady state in the thermodynamic limit.
The reduced fluctuations in the delocalized phase are consistent with the increased mixing allowed by longer range jump operators, as particles are distributed across the system, causing each site to equilibrate with each other.

\begin{figure}
    \centering
    \includegraphics[width=\columnwidth]{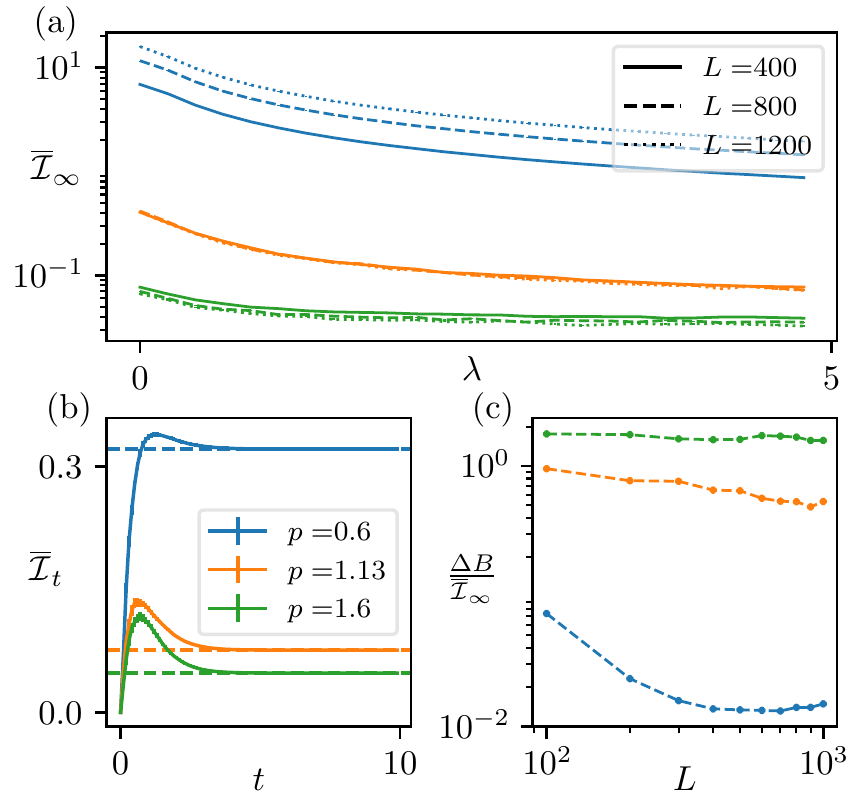}
    \caption{
    (a) NESS bipartite mutual information  $\overline{\mathcal{I}}_\infty$ against hopping strength $\lambda$ for different dissipation and system sizes.  
    (b) Dynamics of the bipartite mutual information $\overline{\mathcal{I}}_t$ between $A$ and $B$ where $L_A=L_B=L/2$ and $L_C=0$, of an initially uncorrelated state, i.e. $\Omega(0)=\textrm{diag}\{a_{1},...,a_{L}\}$, for random real $\{a_n\}$ with system size $L=100$ and $\lambda = 5$. \highlightchange{Dashed lines here denote their associated $\overline{\mathcal{I}}_\infty$.}
    (c) Relative height of the bump in $\overline{\mathcal{I}}_t$ in \highlightchange{(b)} with respect to its $\overline{\mathcal{I}}_\infty$ scaled by $1/\overline{\mathcal{I}}_\infty$, \highlightchange{denoted by} $\Delta B/\overline{\mathcal{I}}_\infty$, against system size, $L$.  
    All curves were averaged over disorder realizations, and error bars are not visible.}
    \label{fig:dynamics}
\end{figure}
\textbf{\textit{Effect of coherent hopping.---}}
To study the influence of a coherent dynamics on the steady state and dynamics, we introduce a nearest neighbor hopping Hamiltonian \highlightchange{through $h$ with $h_{nm} = \lambda$ if $m = n\pm1$ and $0$ otherwise}.
\removeblock{\highlightchange{
\begin{equation}\label{eq: hopping}
    \cancel{H =  \lambda \sum_n \left(c^\dagger_n c_{n+1} + c^\dagger_{n+1} c_{n}\right).}
\end{equation}
}}
Solving Eq.~\ref{eq:correlator-matrix-diff-eq-simplified} in the \highlightchange{eigen}basis of a Hamiltonian, we arrive at the more general solution
\begin{equation}\label{eq: general solution to omega}
\begin{aligned}
    \tilde{\Omega}(t)_{nm} &=  \left( \tilde{\Omega}(0)_{nm}-\frac{i\tilde{\Gamma}_{nm}}{\Delta E_{nm}}\right)e^{-\frac{t}{\tau}+i(E_n-E_m)t} + \frac{i\tilde{\Gamma}_{nm}}{\Delta E_{nm}},
\end{aligned}
\end{equation}
where $E_n$ are the eigenvalues of $h$, $\Delta E_{nm} \doteq E_n-E_m+i$, and $\widetilde{O}_{nm}$ denotes the matrix elements of $O$ in $h$'s basis.
\highlightchange{The timescale ($\tau=1$) corresponds to the real part of the Lindbladian spectrum, originating from $h=0$. Here, creation and annihilation operators for the fermionic eigenmodes each contribute $-1/2$ to the eigenvalue. Since this equation describes the evolution of quadratic gaussian operators, these eigenmodes appear in pairs whose eigenvalues sum to $-1$. For details see SM2. This implies that the timescale is independent of system size, $p$, and details of the dissipation.}
\highlightdelete{The timescale ($\tau$) corresponding to the real part of the spectrum observed in this equation is one, which can be understood through the exact spectrum of the dissipative term in our Lindbladian. 
The jump operators for the creation and annihilation of the fermionic eigenmodes form pairs whose eigenvalues sum to one turning the real part of the spectrum solvable, as detailed in the SM2. This also means that the timescale is independent of system size and details of the dissipation.}

In short, we see that turning on hopping  does not affect the qualitative behavior of the model.
Fig.~\ref{fig:dynamics}\textcolor{blue}{(a)} shows the $\overline{\mathcal{I}}_\infty$ between $A$ and $B$ in the steady state, $L_A=L_B=L/2$ and $L_C=0$: the model still conserves the L dependence of the two phases, with the same critical point, although the $\overline{\mathcal{I}}_\infty$ decreases in magnitude as $\lambda$ is increased for all values of p. However, the critical point doesn't change with $\lambda$. 
Turning our attention on the system dynamics using \eqref{eq: general solution to omega}, the evolution averaged over disorder $\overline{\mathcal{I}}_t$ develops a bump as $\lambda$ is increased from $0$, as shown in Fig.~\ref{fig:dynamics}\textcolor{blue}{(b)}. 
The height of the bump, $\Delta B$, relative to $\overline{\mathcal{I}}_\infty$ at different values of $p$ against $L$ is plotted in Fig.~\ref{fig:dynamics}\textcolor{blue}{(c)}. 
As $L$ is increased, $\Delta B/\overline{\mathcal{I}}_\infty$ decreases for all $p$, but far more rapidly in the delocalized phase. The weak entangling nature of hopping introduces a short time increment in $\overline{\mathcal{I}}_t$ which is subsequently destroyed by the dissipation.

\noindent \textbf{\textit{Conclusions.---}} 
In this work we explored the phenomena of localization in the stationary states of open quantum systems, using quadratic fermionic Lindbladians as a testbed. 
We observed a phase transition from delocalized to localized phases in a model with disordered, long-range bath interactions as the range of the bath interactions is decreased.
We found in the delocalized phase extensive scaling of both mutual information and the participation ratio. In contrast, as the localized phase is approached $\overline{\mathcal{I}}_\infty$ tends to zero and the participation ratio is $O(1)$ in the short-range limit of bath interactions.
Noting a slow decrease of participation ratio close to the phase transition, we further considered measures of multifractality, finding observations close to the transition consistent with systems exhibiting such a phenomena.
In the delocalized phase, we show that the steady state occupation is homogeneous  while the localized phase exhibits a significant heterogeneity which survives in the thermodynamic limit.
Finally, we considered the effect of a coherent evolution on the stationary state and dynamics, finding that the critical point remain unchanged by hopping terms in the Hamiltonian.

Our work realizes the effects of localization in steady states of open quantum systems and shows the significance of conventional Anderson localization in dissipative systems. \highlightchange{Stability of this phenomena in the presence of} interactions could have relevance to questions of many-body localization. Furthermore, it would be interesting to understand this phenomenon in the context of measurement induced entanglement transitions in quantum trajectories.

\textbf{\textit{Acknowledgments.---}}
We would like to thank Christopher Turner, Dawid Paszko and Marcin Szyniszewski for their valuable insights and discussions. 
A.P.\ is funded by the European Research Council (ERC) under the EU’s Horizon 2020 research and innovation program via Grant Agreement No.~853368.
The authors acknowledge the use of the UCL Myriad High Performance Computing Facility (Myriad@UCL), and associated support services, in the completion of this work.

\bibliography{disorderedopenfreefermions}

\input{supplementary_material.tex}

\end{document}

%% file: supplementary_material.tex




\preprint{APS/123-QED}

\title{Supplementary materials}

\author{Alejandro Cros Carrillo de Albornoz}
\affiliation{Department of Physics and Astronomy, University College London, Gower Street, London WC1E 6BT, United Kingdom}

\author{Dominic C. Rose}
\affiliation{Department of Physics and Astronomy, University College London, Gower Street, London WC1E 6BT, United Kingdom}

\author{Arijeet Pal}
\affiliation{Department of Physics and Astronomy, University College London, Gower Street, London WC1E 6BT, United Kingdom}

\maketitle
\subsection*{1. Mathematical details}\label{appendix: Lindbland eq}
Our model begins from a general quadratic Fermionic Lindbladian without superconductive terms in the Hamiltonian.
\begin{align}\label{eq:lindbland-eq}
\frac{d}{dt} \rho(t) =
& -i\left[\sum_{n,m}h_{nm}c^\dagger_n c_m,\rho(t)\right]\nonumber\\
&+ \sum_{jk}\Gamma_{jk} \left(c_k^\dagger\rho(t) c_j-\frac{1}{2}\{c_j c_k^\dagger,\rho(t)\}\right)  \nonumber\\
&+ \sum_{jk} B_{jk} \left(c_j\rho(t) c^\dagger_k-\frac{1}{2}\{c_k^\dagger c_j,\rho(t)\}\right)  \;, 
\end{align}
from which a closed equation for the time evolution of the two point correlation matrix $\Omega_{jk} = \textrm{Tr}(\rho c^\dagger_j c_k)$ can be derived \highlightchange{by multiplying \eqref{eq:lindbland-eq} by $c^\dagger_j c_k$ and then taking the trace \cite{Mahajan2016}, after some algebra we obtain}
\begin{equation}\label{eq:correlator-matrix-diff-eq}
    \frac{d}{dt} \Omega(t) = i[h^T,\Omega(t)]+\frac{1}{2}\{\Gamma,\mathbb{I}-\Omega(t)\}-\frac{1}{2}\{B,\Omega\}.
\end{equation}
This equation is guaranteed to keep the correlation matrix physical due to constraints inherited from the Lindblad equation: hermiticity is ensured by the hermiticity of $\Gamma$, $B$ and $h$; positivity is ensured by the positivity of $\Gamma$ and $B$.
Physically, $\Gamma$ and $B$ correspond to competing bath interactions which respectively add and remove particles from the system. Existing in isolation, $\Gamma$ would push the system into a state where every site is occupied by a particle, while $B$ would lead to only holes.

We specifically focus on the case when $[\Gamma,B]=0$. 
The eigenvalues $\{\alpha_n\} $ of $\Gamma$ and $\{\beta_n\}$ of $B$ and their corresponding eigenvectors $\{\ket{g_n}\}$ fully describe the behavior of the dissipation. 
In the case when $h=0$, the eigenstates may be thought of as describing quasiparticle modes of the system: in the Lindbladian description, these eigenstates may be used to diagonalize the dissipative term, with each state resulting in a potentially delocalized jump operator $d_m=\sum_i \left\langle i | g_m \right\rangle c_i$.
Since we assume $\Gamma$ and $B$ commute, these jump operators appear in conjugate pairs which add and remove excitations from these modes at rates determined by the corresponding eigenvalues.
The likelihood of these modes being occupied is therefore described by the ratios $\alpha^n/\beta^n\in (0,\infty)$, which may be thought of as representing the relationship between the energy of each mode and the temperature of the bath is is coupled to.

Finally, we further specialize to the case when $B=\mathbb{I}-\Gamma$, simplifying Eq.~\eqref{eq:correlator-matrix-diff-eq} to
\begin{align}\label{eq:app-correlator-matrix-diff-eq-simplified}
    \frac{d}{dt} \Omega(t) =  i[h^T,\Omega(t)] + \Gamma - \Omega(t).
\end{align}
In the spectral basis of $h$ the equations for each component decouple
\begin{align}
    \frac{d}{dt} \widetilde{\Omega(t)}_{nm} =  i(E_n-E_m+i)\widetilde{\Omega(t)}_{nm}+ \widetilde{\Gamma}_{nm},
\end{align} 
where $E_n$ are the eigenvalues of $h$ and $\widetilde{O}_{nm}$ denotes the matrix elements of $O$ in $h$'s basis. 
The solution to Eq.~\eqref{eq:app-correlator-matrix-diff-eq-simplified} is then 
\begin{equation}\label{eq:general-solution-to-omega-appendix}
\begin{aligned}
    \widetilde{\Omega(t)}_{nm} &= e^{-t} \left( \widetilde{\Omega(0)}_{nm}-\frac{i\widetilde{\Gamma}_{nm}}{\Delta E_{nm}}\right)e^{i(E_n-E_m)t} + \frac{i\widetilde{\Gamma}_{nm}}{\Delta E_{nm}},
\end{aligned}
\end{equation}
where $\Delta E_{nm} \doteq E_n-E_m+i$.
Note that the steady state is independent of the initial conditions, indicating it is unique.
A key benefit of the above is that it is numerically efficient to construct its time evolution and its stationary state up to and beyond a system size of $10^4$ sites. 

In the absence of $H$, Eq.~\eqref{eq:general-solution-to-omega-appendix} reads 
\begin{equation}
    \Omega_{nm}(t) = \Gamma_{nm}(1-e^{-t})+\Omega_{nm}(0)e^{-t},
\end{equation}
in the original basis of $\Gamma$

\subsection*{2. Exact solution without Hamiltonian term}
In the case when $h=0$, an exact solution can be accessed by first separating the system into a set of independent fermionic modes based on the spectrum of $\Gamma$, then solving each independent system, as follows.
For generality we show the result here for when $[\Gamma,B]=0$, rather than any specific $B$ which satisfies this equation.
Writing the spectrum of $\Gamma$ as $\Gamma\ket{g_i}=\gamma_i\ket{g_i}$, and the eigenvalues of $B$ as $b_i$, we may rewrite our Lindbladian as
\begin{align}
    \mathcal{L}(\rho)=\sum_m
    &\left[\gamma_m d_m^\dagger \rho d_m - \frac{\gamma_m}{2}\left\{\rho,d_m d_m^\dagger\right\}\right.\nonumber\\
    &\left.b_m d_m^\dagger \rho d_m - \frac{b_m}{2}\left\{\rho,d_m d_m^\dagger\right\}\right]\nonumber\\
    =\sum_m
    &\mathcal{L}_m(\rho),
\end{align}
where $d_m=\sum_i \left\langle i | g_m \right\rangle c_i$.
Each term in this sum corresponds to an independent $2$-level fermionic mode evolving according to a $4\times4$ Lindbladian.

For a given independent subsystem, with basis $\ket{0_m}$, $\ket{1_m}$ such that $d_m^\dagger\ket{0_m}=\ket{1_m}$ and $d_m\ket{1_m}=\ket{0_m}$, we calculate the matrix elements $\tr\left[\ket{i_m}\bra{j_m}\mathcal{L}_m\left(\ket{i'_m}\bra{j'_m}\right)\right]$.
Ordering the basis as $\ket{1_m}\bra{1_m}$, $\ket{0_m}\bra{0_m}$, $\ket{0_m}\bra{1_m}$, $\ket{1_m}\bra{0_m}$ we find
\begin{align}
    \mathcal{L}_m=
    \begin{pmatrix}
        -b_m & \gamma_m & 0 & 0 \\
        b_m & -\gamma_m & 0 & 0 \\
        0 & 0 & -\frac{\gamma_m+b_m}{2} & 0 \\
        0 & 0 & 0 & -\frac{\gamma_m+b_m}{2}
    \end{pmatrix},
\end{align}
finding that the coherences are already eigenmodes, while the occupation expectations support a $2\times2$ block.
This may be diagonalized to find eigenvalues of $0$ and $-\gamma_m-b_m$, with corresponding left and right eigenvectors.
Eigenmodes of the full Lindbladian can then be constructed by taking tensor products of the eigenmodes of each independent fermionic mode, and rewriting the state in terms of the original position-space creating and annihilation operators, with their corresponding eigenvalues given by the sum of the eigenvalues for each subsystem eigenmode used in the product.

In the case where $b_m=1-\gamma_m$, as we have in the main text, this result explains why our correlation matrix evolution equation has a uniform relaxation rate of $1$, independent of system size and $p$.
Since in this case the eigenvalues of each single-fermion Lindbladian are either $0$, $-1/2$ or $-1$, eigenvalues of the full Lindbladian must be multiples of $-1/2$.
Considering the space of quadratic states, there are three classes of terms which contribute: $d_m^\dagger d_m=\ket{1_m}\bra{1_m}$, $d_m d_m^\dagger=\ket{0_m}\bra{0_m}$, $d_m^\dagger d_n=\ket{1_m}\bra{0_m}\otimes\ket{0_n}\bra{1_n}$.
The first two have support only on single-fermion eigenmodes with eigenvalues of $0$ and $-1$, while the third consists of a tensor product of two single-fermion eigenmodes each with eigenvalue $-1/2$, and thus has an overall eigenvalues of $-1$.
As such, all quadratic states reside in a vector subspace with support on eigenmodes with eigenvalues of $0$ and $-1$.
Since gaussian states are a subspace of such matrices, they reside in the same vector subspace, and therefore possess the same uniform relaxation time of $1$, also imparted on the evolution of their correlation matrices.
We therefore see that this relaxation time, and its parameter independence, has its origin in the precise relationship we chose in taking $B=I-\Gamma$, leading to a particular balancing of transition rates in the classical stochastic evolution each independent fermionic mode undergoes.

\subsection*{3. Multifractality}
\begin{figure}
    \centering
    \includegraphics[width=\columnwidth]{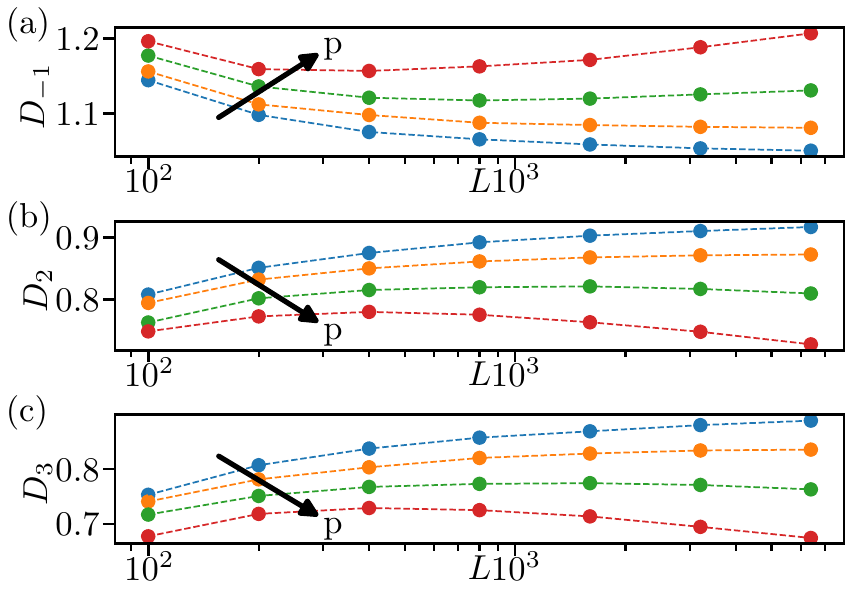}
    \caption{(a-c) Generalized fractal dimensions as a function of system size for $p=[0.95, 1.0, 1.05, 1.1]$, increasing along the arrow.}
    \label{fig:multifractality}
\end{figure}

To further understand the intermediate region between phases, particularly notable in the participation ratio, we calculate the generalized fractal dimensions encoding the scaling behavior of moments, as considered in other works on PBRMs \cite{Cuevas2001,Cuevas2003}.
For a given eigenstate $\psi$ viewed as a single-particle wavefunction in a 1D system with $L$ sites, we denote the probability of the particle being found in a box from $i$ to $i+l-1$ as
\begin{align}
    p_l(i)=\sum_{j=i}^{i+l-1}|\psi_j|^2.
\end{align}
The $q$th moment of this probability distribution over disjoint boxes of length $l$ is given by
\begin{align}
    \chi_q(l,L)=\sum_{i=0}^{L/l} p_l^q(il).
\end{align}
Finally, the generalized fractal dimensions are then extracted from the scaling behavior of these moments.
Assuming the moments satisfy a large deviation principle
\begin{align}
    \chi_q(l,L)\propto\left(\frac{l}{L}\right)^{D_q},
\end{align}
we have
\begin{align}
    D_q=\lim_{\delta\rightarrow0}\frac{\ln(\chi_q(\delta L, L))}{\ln(\delta)},
\end{align}
where $\delta=l/L$ is the fraction of the system contained in each region.
Fig.~\ref{fig:multifractality} shows how estimates of these change as $L$ increase for a variety of $q$ and $p$.
We note two properties of this data consistent with multifractal phenomena \cite{Cuevas2001,Cuevas2003,Mace2019}.
Firstly, in the vicinity of $p=1.0$ the dimensions become approximately constant at large $L$, diverging away from this constant with $L$ for $p$ further from the critical point.
When these exponents become independent of system size for all moments, the statistical properties of the particles position depend only on the size of the subsystem relative to the total system size, not the absolute subsystem size.
Secondly, the fractal dimension attains noticeably distinct values as $q$ is varied, in contrast to non- and mono-fractal systems in which the fractal dimension remains largely constant.

\subsection*{4. $\overline{\textrm{PR}}$ fit}
\begin{figure}
    \centering
    \includegraphics[width=\columnwidth]{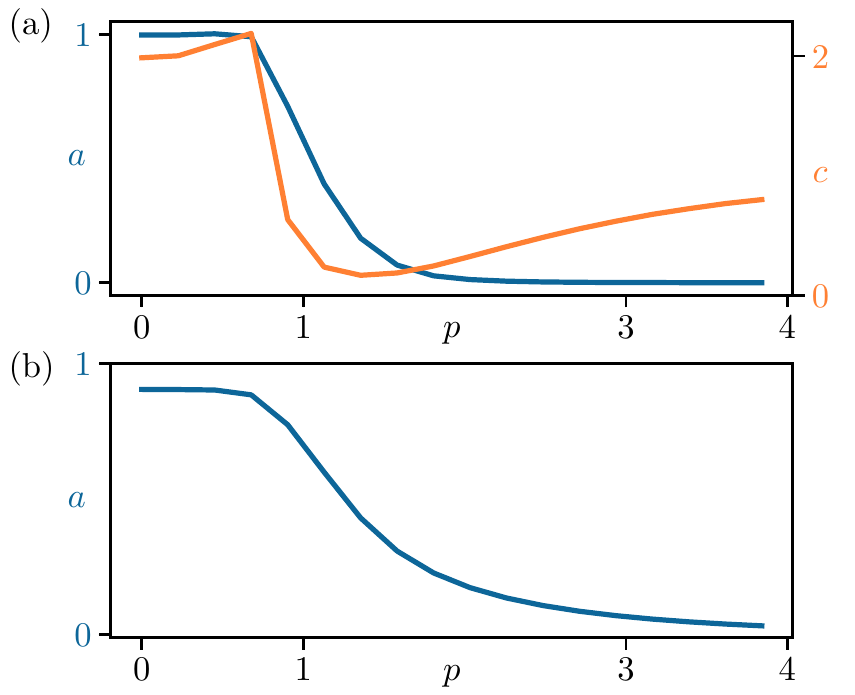}
    \caption{Coefficients as a function of $p$ found when fitting $\overline{\mathrm{PR}}(L,p)=L^{a(p)}/c(p)$ to $\overline{\textrm{PR}}(L)$ curves.}
    \label{fig:pr_fit}
\end{figure}

In the main text, the fitting function $\overline{\mathrm{PR}}(L,p)=L^{a(p)}/c(p)$ is used to study the scaling behavior of the $\overline{\mathrm{PR}}$ contains two parameters: the exponent $a$ mentioned in the main text, and an overall scaling coefficient $c$.
For completeness, in Fig.~\ref{fig:pr_fit}\textcolor{blue}{(a)} we show the $p$ dependence for both these coefficients.
For comparison, in Fig.~\ref{fig:pr_fit}\textcolor{blue}{(b)} we also show the exponent $a(p)$ found when fitting $c(p)=1$ for all $p$: the lack of $c(p)$ to allow scaling to increase the overall $\overline{\mathrm{PR}}$ produced by the fit, a key to accurately fitting in the intermediate regime, results in a slower decay of the exponent as $p$ is increased.
However, the overall behavior is qualitatively the same.

\subsection*{5. Numerical properties of $\Gamma$}
\subsubsection*{5.1. Maximum eigenvalue statistics}
\begin{figure}[H]
    \centering
    \includegraphics[width=\columnwidth]{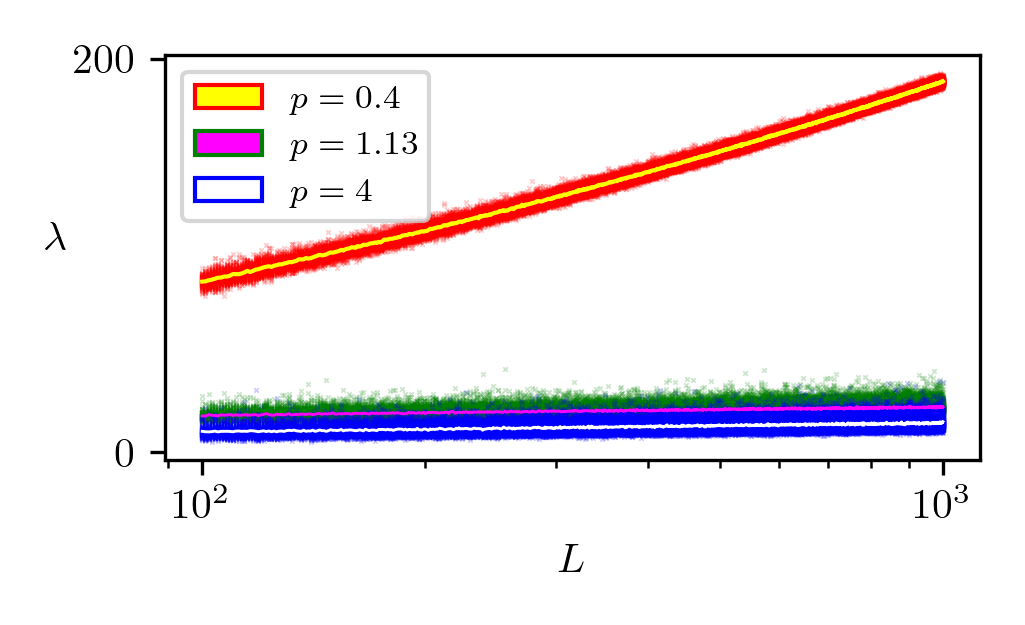}
    \caption{Eigenvalues of $X^\dagger X$ (markers) and their average against $L$ (lines).}
    \label{fig: max_lamb}
\end{figure}

The behavior of the largest eigenvalue of $X^\dagger X$, $\lambda_\text{max}$, is of great importance to the dynamical study of $\rho$ because 
\begin{equation}
\Gamma \doteq X^\dagger X/2 \lambda_\text{max},
\end{equation}
as previously defined in the main text (which again this choice tamed the semi-positiveness of $\rho$). 
This maps the spectrum of $\Gamma$ to the $[0,1]$ interval regardless of $p$ or $L$.
Thus, it is worth studying the behavior of $\lambda_{\text{max}}$ used for this definition of $\Gamma$ for different values of $p$ and $L$, Figure \ref{fig: max_lamb}. 
One finds that in the localized phase $\lambda_{\text{max}}$ remains roughly fixed at a constant value for all such $p$ and $L$ whereas in the thermal phase $\lambda_{\text{max}}$ grows as a power-law with $L$.
\subsubsection*{5.2. Power-law decay of $\Gamma$ matrix elements}
\begin{figure}[H]
    \centering
    \includegraphics[width=\columnwidth]{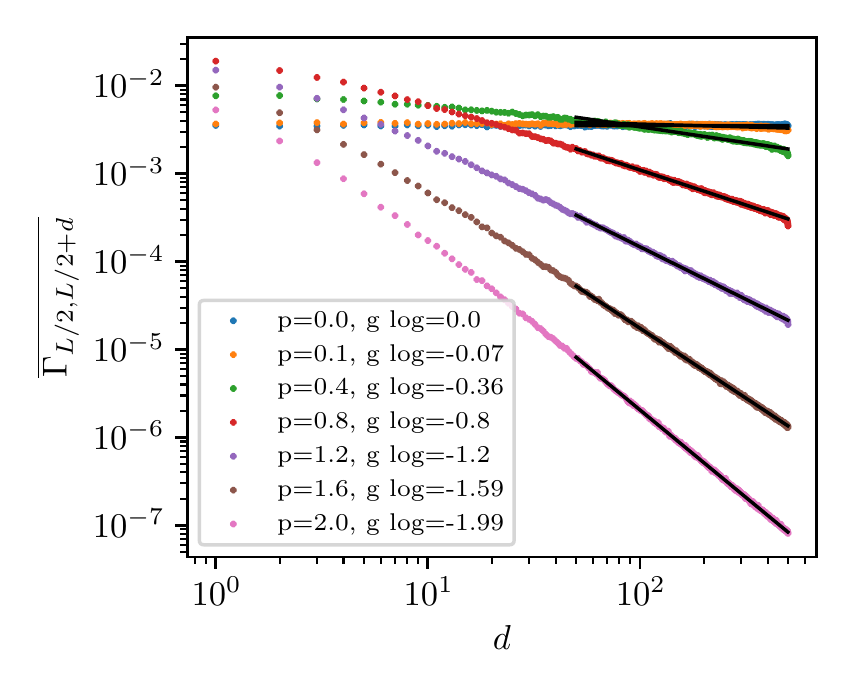}
    \caption{Log-log plot of the decay of the matrix elements of $\Gamma$ as a distance $d$ away from the diagonal. For large enough $d$, the decay is given by a power law with an exponent approximately equal to $p$. In the legend, $g$ $\log$, is the linear gradient of the black lines fitted to the log-log data. Note how $p\sim g$ $\log$, suggesting a power law decay with power $p$ for large enough $d$.}
    \label{fig: Gamma_vs_p}
\end{figure}

$\Gamma$'s matrix elements decay away from the diagonal as the distance to it, $d$, is increased. This decay is intimately related to $p$ via
\begin{equation}\label{eq: the model}
    X_{jk} = \frac{x_{jk}}{(|j-k|+1)^p}.
\end{equation}
As depicted in Figure \ref{fig: Gamma_vs_p}, for large enough $d$, the decay of the matrix elements of $\Gamma$ away from the diagonal is roughly given by
\begin{equation}
    ||\Gamma_{L/2, L/2+d}|| \sim \frac{1}{d^p}  \quad \text{for} \quad 1\ll d.
\end{equation}
As $\Gamma$ is hermitian, this observation connects the localization transition of $\Gamma$'s eigenvectors to the already mentioned power-banded localized Hamiltonian models in the literature. 

\subsubsection*{5.3. Power-law decay of $\Gamma$ eigenvectors}
\begin{figure}[H]
    \centering
    \includegraphics[width=\columnwidth]{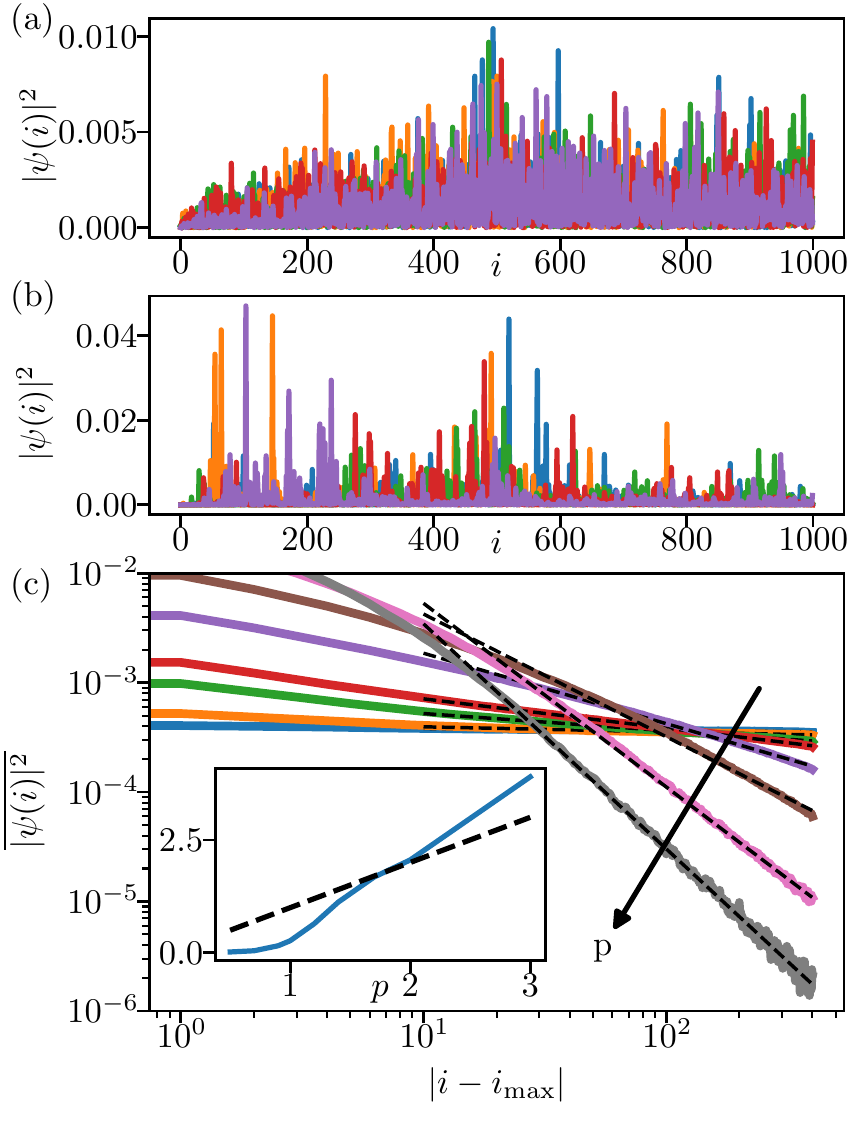}
    \caption{
        (a-b) Five sample eigenstates with a maximum magnitude value within the middle 25th of the system at (a) $p=0.5$ and (b) $p=1.1$.
        (c) Log-log plot of the averaged positional probability as a function of distance from the most probably site, for $p=[0.5, 0.7, 0.9, 1.0, 1.2, 1.4, 1.7, 2.0]$. 
        Averaged over eigenstates with most probable sites located in the middle 5th of the system, and averaged over $1000$ disorder realizations.
        Linear fits of the probabilities for distances of $100$ to $1000$ are shown by the dashed lines.
        Inset: (solid) gradients of the linear fits are shown vs $p$, (dashed) the curve $y=p$.}
    \label{fig:wavefunction_decay}
\end{figure}

Prior studies of PRBMs have observed power-law decay of eigenstates away from a central site, particularly closer to the transition within the localized phase.
This is in contrast to eigenstates in short-range models exhibiting exponential decay.
In Fig.~\ref{fig:wavefunction_decay} we present data suggesting our PRBM model exhibits the same phenomena, with small deviations from power-law behavior at lower values of $p$.

\subsection*{6. Mutual information with boundaries}
\begin{figure}[H]
    \centering
    \includegraphics[width=\columnwidth]{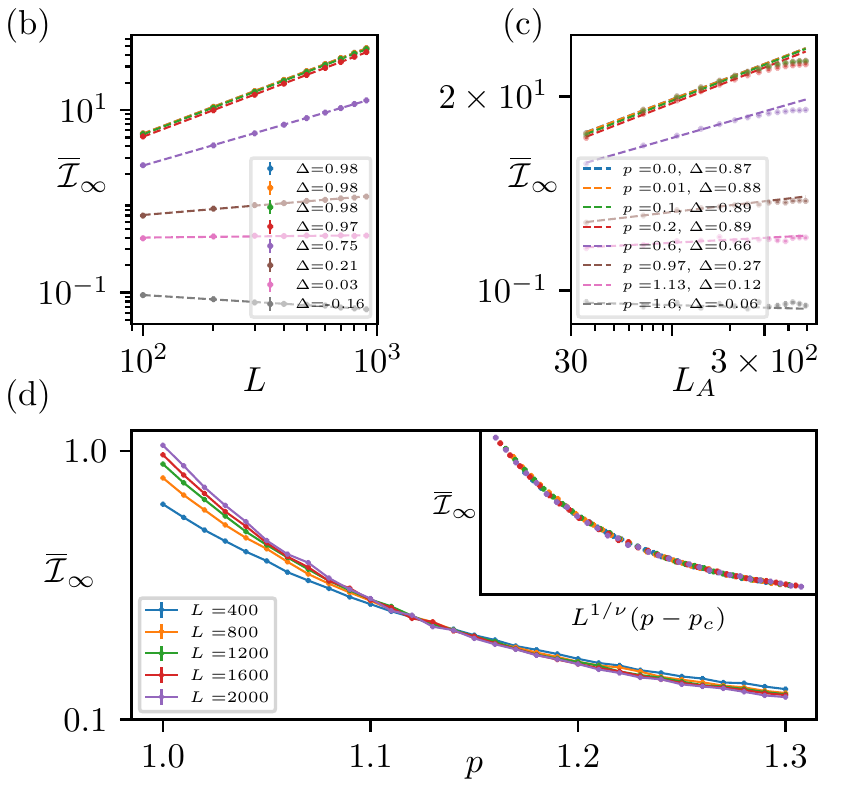}
    \caption{Steady state mutual information between $A$ and $B$ including their boundary term $\partial_{AB}$.}
    \label{fig: MI fig with boundaries}
\end{figure}

When we set to explore the mutual information behavior between two subsystems, one needs to choose whether or not to include the boundaries that separate these two subsystems. 
In the main text, we decided to exclude the boundaries between $A$ and $B$ so when $L \rightarrow \infty$ even if $A/L$ and $B/L$ are set constant the boundary $\partial_{AB}$ contribution to $\overline{\mathcal{I}}_\infty$ dies off as $\partial_{AB}$ grows. 
This is because it filters the very short-range entanglement contribution of $\partial_{AB}$ for sufficiently large $L$. 
This can be seen in the area-law phase in Fig. 1 of the main text. 
Here we redo our analysis of the mutual information without removing $\partial_{AB}$ in Fig. \ref{fig: MI fig with boundaries}. 
As expected, $p_c$ shifts into the area law phase as now the short-range entanglement is a more dominant contribution to $\overline{\mathcal{I}}_\infty$. 
This also illustrates the strength of this contribution towards $\overline{\mathcal{I}}_\infty$, which can be seen in this shift and the overall substantially larger magnitude of $\overline{\mathcal{I}}_\infty$.
Note that in contrast to the unusual $L_A$ dependence we observe in the case without boundary contributions, here we see a clear change from a sub-extensive power-law dependence at low $p$ to an area law dependence at high $p$.

\subsection*{7. Numerical details}
\subsubsection*{7.1. Disorder realizations}
In Fig. 1 of the main text (b), disorder realizations are linearly decreased from $500$ at $L=100$ to $100$ at $L=1000$. In (c), $20$ disorder realizations were used. Fig. 2 of the main text uses identical disorder realizations to Fig. 1. In Fig. 3 of the main text, from $L=400$ to $L=2000$, disorder realizations were $[500,200,200,200,100]$. The inset of Fig. 3 of the main text was done using identical numbers of disorder realizations. 

Fig. 5 of the main text was computed by randomly generating $\Gamma$ at each $p$, $10^3$ times, and from this obtaining the temperature distribution shown. The inset was computed at $L=[50,100,\cdots,1000]$ and averaged with disorder realizations $[2000, 1900, \cdots, 100]$ respectively for each value of $p$.

Fig. \ref{fig: max_lamb} was computed by randomly sampling $X^\dagger X$ in the $L=[10^2,10^3]$ interval $10^5$ times for each value of $p$ shown.

Fig. \ref{fig: Gamma_vs_p} was computed at $L=10^3$ and averaged over $10^3$ realizations.

\subsubsection*{7.2. Finite-size scaling}
To perform a finite-size scaling analysis on the MI data, we optimize a standard loss function which measures the spread of the data, see e.g. Ref. \cite{zabalo}.
Given our data possesses non-zero error bars from averaging over disorder realizations, we perform this analysis by sampling gaussian perturbations of our dataset, sampling noise for each data point with a standard deviation equal to the error in the mean for that data point.
For each sample of a noise-perturbed dataset, we then optimize the loss to find a corresponding critical point and exponent.
The expected value and variance of these resulting parameters are then calculated, using $1000$ noisy realizations of our dataset.
